\documentclass[11pt,twoside]{article}

\usepackage{asp2006}
\usepackage{epsf}
\usepackage{psfig}
\usepackage{lscape}

\begin{document}
\title{What is a galaxy? How Cold is Cold Dark Matter? Recent progress
in Near Field Cosmology}   

\author{Gerard Gilmore$^1$, Dan Zucker$^1$, Mark Wilkinson$^2$,
  Rosemary F.G. Wyse$^3$, Vasily Belokurov$^1$, Jan Kleyna$^4$,
  Andreas Koch$^5$, N. Wyn Evans$^1$, Eva K. Grebel$^6$}
{\footnotesize{ 1) Institute of Astronomy, Cambridge, UK; 2) Leicester
  University, UK; 3) Johns Hopkins University, Baltimore, USA; 4)
  Institute for Astronomy, University of Hawaii, USA; 5) Dept of
  Physics and Astronomy, UCLA, USA; 6) Astronomisches Rechen-Institut,
  Zentrum f\"ur Astronomie der Universit\"at Heidelberg, Germany. }}

\begin{abstract}
  There has been a vast recent improvement in photometric and
  kinematic data for star clusters, Ultra Compact dwarfs, galactic
  nuclei, and local dSph galaxies, with Subaru contributing
  substantially to the photometric studies in particular. These data
  show that there is a bimodal distribution in half-light radii, with
  stable star clusters always being smaller than 35pc, while stable
  galaxies are always larger than 120pc.  We extend the previously
  known observational relationships and interpret them in terms of a
  more fundamental pair of intrinsic properties of dark matter itself:
  dark matter forms cored mass distributions, with a core scale length
  of greater than about 100pc, and always has a maximum central mass
  density with a narrow range. The dark matter in dSph galaxies
  appears to be clustered such that there is a mean volume mass
  density within the stellar distribution which has the very low value
  of about 0.1$M_{\odot}$ pc$^{-3}$.  None of the dSphs displays
  kinematics which require the presence of an inner cusp, while in two
  dSphs there is evidence that the density profile is shallow (cored)
  in the inner regions. The maximum central dark matter density
  derived is model dependent, but is likely to have a mean value
  (averaged over a volume of radius $10$pc) of about 0.1$M_{\odot}$
  pc$^{-3}$, which is 5GeV/c$^2$cm$^{-3}$). Galaxies are embedded in
  dark matter halos with these properties; smaller systems containing
  dark matter are not observed.
\end{abstract}

\section{What is a galaxy}   

The existence of a clear observational distinction between massive
star clusters and low-mass galaxies has been substantially
strengthened recently, both through detailed studies of more luminous
and massive star clusters in a wide range of environments, and through
discovery of a large number of extremely low-luminosity satellite
galaxies around the Milky Way (e.g.~Willman et al.~2005; Belokurov et
al.~2007)and around M31, mostly based on imaging from the SDSS,
usually with supplementary - and crucial! -- Subaru studies.

\begin{figure}[!ht]
\plotfiddle{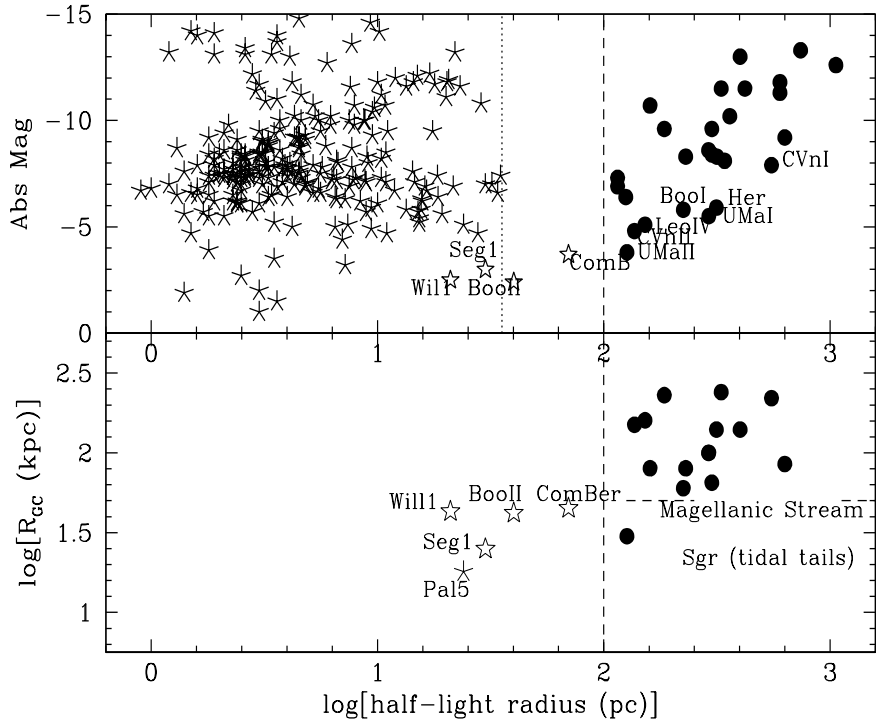}{10truecm}{-90}{45}{50}{-180}{300}
\caption{Top: Absolute magnitude ${\rm M_V}$ {vs} (logarithmic)
  half-light radius for well-studied stellar systems. Filled symbols
  are objects classed as galaxies, asterisks are objects classed as
  star clusters of various types, open stars are newly discovered
  objects of uncertain nature. Lower: (log) Galactocentric distance vs
  (logarithmic) half-light radius for the dSph and uncertain objects,
  and the (tidal-tail) star cluster Pal~5. All the uncertain objects
  are in a region where Galactic tides must be important. There is no
  known stable system with half-light radius between 35pc and 120pc.
}
\end{figure}

Fig.~1 shows the current sample in a plot of half-light radius against
absolute magnitude in the V-band.  Star clusters in all studied
environments, with luminosities over the whole range from M$_V = -4$,
L$\sim 10^3$~L$_\odot$, up to M$_V=-15$, L$\sim10^9\, L_\odot$, and a
wide range of ages, invariably have characteristic scale sizes $r_{h}$
less than about 30pc.  Thus the dynamical range over which both star
clusters and galaxies exist, and over which there is a distinct size
dichotomy, has now been established to cover some six orders of
magnitude in stellar luminosity. Figure~1 makes evident that there is
a robust maximum radius to star clusters, at all luminosities.
Figure~1 also illustrates that the dSph galaxies in both the Milky Way
and in M31 have a minimum characteristic radius, and this minimum is a
factor of $>4$ larger than the largest star clusters.  With the
exception of the recently discovered object ComaBer (Belokurov et
al.~2007) -- the largest and brightest of the four systems indicated
by open stars in Fig.~1 -- there is no known object in the size-gap
between $\sim35$pc and $\sim120$pc. ComaBer manifestly merits further
study, but is one of the new dSph which lie between the Sgr dwarf and
the Magellanic Clouds, and which show significant indications of tidal
disruption.

It is more correct to say that, modulo ComBer, there is no known
stable object in the size gap.  Fig~1 (lower panel) illustrates this
further, showing that all possible transition objects are in a
tidally-affected region, close to the Galactic centre. This
intermediate size can be occupied transiently by a larger object (a
dwarf galaxy) in the very late stages of disruption by external
(Galactic) tides, or by a small object (globular cluster) in the last
stages of evaporation. In the first of these cases a low velocity
dispersion compact core can be generated transiently, if outer,
hotter, stars are removed by a suitable tide. In the second case
the density profile changes systematically from the small scale
typical of a compact star cluster to a very large scale, almost
constant density, covering all possible radii during that
(short-lived) process.  Segue~1, Will~1 and Boo-II are also newly
discovered and interesting tests of the conclusions of this section,
with all showing photometric evidence for significant tidal
disturbance. The two largest Galactic globular clusters are Pal~14,
with size 28~pc, and Pal~5, 24~pc. Pal~5 is in an advanced stage of
tidal disruption, with prominent streams of stripped stars
(\cite{Od2003}).

The largest Ultra Compact Dwarf galaxies (UCDs), also with size 25pc,
are associated with the centre of the Virgo cluster, and the galaxy
M87, and were for a long time suspected of being small galaxies
severely affected by tides. Our interpretation of their status, based
on their position in Figure~1, is that they are simply very massive
star clusters, with no associated dark matter, perhaps embedded in a
lower surface brightness galaxy which has an associated dark matter halo.
There are two recent detailed dynamical
analyses of the masses of ultra compact dwarf galaxies, by
\citet{Hilker07} and by \citet{EGDH07}. They show that all these
systems are similar, in structure and dynamics, and that the dynamical
mass-to-light ratios for the UCDs are (almost) consistent with simple
stellar models: there is no evidence for any dark matter associated
with these objects.  They are the (very)
high-mass/high-luminosity extreme of more typical globular cluster
populations. We note a recent study by \citet{DHK08} which has 2 UCD
and one compact elliptical in or close to the size gap. The compact
elliptical is M32, which is in tidal interaction with M31.  The 2 UCDs
are in fact systems in which there is a small UCD (radius 10pc)
embedded in an extended (200pc) halo. The plotted size is the mean of
these two, and is not the UCD.  The size gap remains empty of stable
objects. We have apparently really identified a minimum size for a
galaxy.

\begin{figure}[!ht]
\plotone{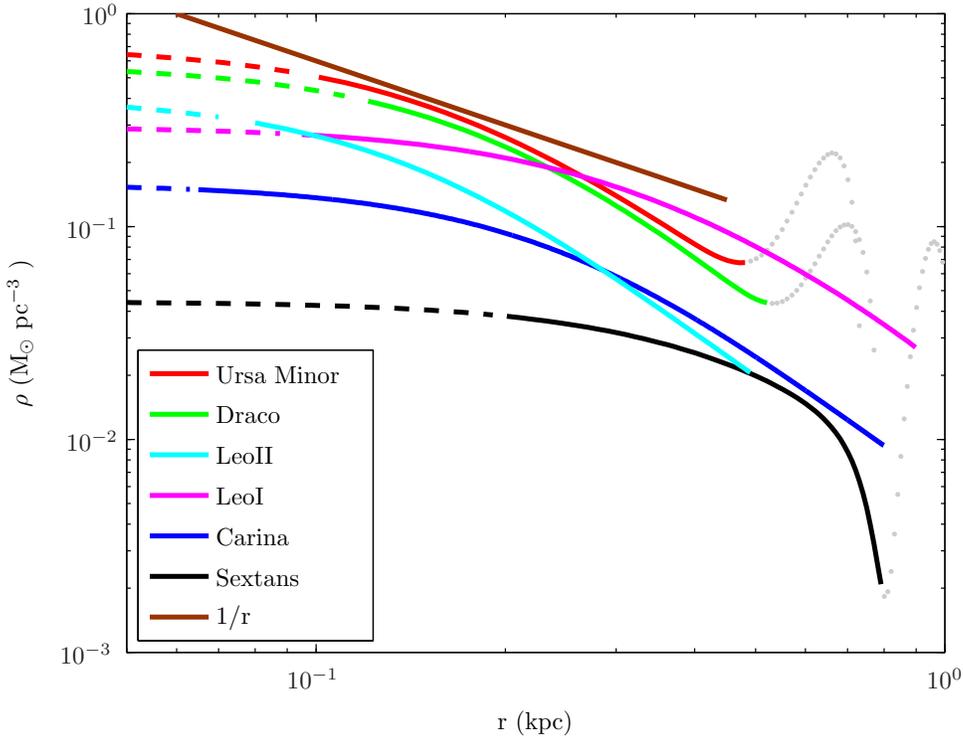}
\caption{Derived inner mass distributions from isotropic Jeans'
  equation analyses for six dSph galaxies. The modelling is reliable
  in each case out to radii of log (r)kpc$\sim0.5$.  The general
  similarity of the inner mass profiles is striking, as is their
  shallow profile, and their similar central mass densities. Also
  shown is an $r^{-1}$ density profile, predicted by many CDM
  numerical simulations. The dynamical analyses are described in
  more detail in \cite{GG07}. }
\end{figure}

\section{Mass profiles in dSph galaxies}

We can know more than just the sizes of the dSph galaxies. Extensive
kinematic studies provide the information to determine dark mass profiles 
in these galaxies. \cite{GG07} describe in detail the current status
of a uniform Jeans'-equation based study. Fig~2 here summarises that
information. We conclude that the Jeans analysis demonstrates 
that the observed velocity dispersion profiles and cored light
distributions of dSphs are consistent with their
inhabiting dark matter haloes with central cores. Where sufficient
data exist, more sophisticated distribution function modelling also,
in every case studied, produces a small preference for cored dark
matter mass distributions at small scales. We find no evidence in any
case to support the cusped profile predicted by standard CDM models.
The implications of this are discussed further by \cite{GG07}, and
summarised in our abstract.

Briefly, it seems that all dSph galaxies have cored mass
distributions, with surprisingly low central mass densities. This mass
distribution is associated with an even more surprising large spatial
scale length. The central mass densities seen are of order 10GeV/c$^2$
cm$^{-3}$, while we note that the mass scale of the Higgs, and its
heavier partners, is above 100 GeV/c$^2$ per particle. The
corresponding length scales are of order $10^{18}$m, significantly
longer than expected for extremely low velocity dispersion
`cold' particles.




\begin{thebibliography}{}
\bibitem[Belokurov et al.(2007)]{catsdogs} Belokurov, V. et al 2007
  \apj, 654, 897

\bibitem[Dabringhausen et al (2008)]{DHK08} Dabringhausen, J., Hilker,
  M., \& Kroupa, P., 2008 arXiv:0802.0703

\bibitem[\protect\citeauthoryear{Evstigneeva, Gregg, Drinkwater, \&
Hilker}{2007}]{EGDH07}Evstigneeva, E.A., Gregg, M.D., Drinkwater,
M.J., \& Hilker, M., 2007 AJ 133 1722

\bibitem[Gilmore et al.(2007)]{GG07} Gilmore, G., Wilkinson, M.,
  Wyse, R.F.G., Kleyna, J., Koch, A., Evans, N. Wyn, \& Grebel, Eva T.,
  2007 ApJ 663 948 

\bibitem[\protect\citeauthoryear{Hilker etal }{2007}]{Hilker07} Hilker,
M., Baumgardt, H., Infante, L., et al. 2007 A\& A 463 119

\bibitem[Odenkirchen et al. (2003)]{Od2003}
Odenkirchen, M., etal 2003 AJ 126 2385 

\bibitem[Willman et al. (2005b)]{Will05} Willman, B., et al.\ 
2005, \aj, 129, 2692 
\end{thebibliography}
\end{document}